\documentclass[lettersize,journal]{IEEEtran}

\IEEEoverridecommandlockouts
\usepackage{cite}
\usepackage{amsmath,amssymb,amsfonts}
\usepackage{algorithmic}
\usepackage{graphicx}
\usepackage{textcomp}
\usepackage{xcolor}
\usepackage{multirow}
\usepackage{subfigure}
\usepackage{indentfirst}
\usepackage{threeparttable}
\usepackage{pifont}
\usepackage{enumerate}
\usepackage{setspace}
\usepackage{bm}
\usepackage{caption}
\usepackage{upgreek}
\usepackage{cases}
\usepackage[switch]{lineno}
\usepackage{textcomp}
\usepackage{algorithm}
\usepackage{cite}
\usepackage{mathrsfs}
\usepackage{float} 
\usepackage{subfigure}  
\usepackage{algorithm}

\def\BibTeX{{\rm B\kern-.05em{\sc i\kern-.025em b}\kern-.08em
    T\kern-.1667em\lower.7ex\hbox{E}\kern-.125emX}}
\usepackage{epstopdf}
\begin{document}

\title{\huge{A New Intelligent Reflecting Surface-Aided
Electromagnetic Stealth Strategy}\\
}

\author{
	Xue Xiong, 
	 \IEEEauthorblockN{Beixiong Zheng,~\IEEEmembership{Senior Member,~IEEE}}, \IEEEauthorblockN{A. Lee Swindlehurst,~\IEEEmembership{Fellow,~IEEE}},\\ \IEEEauthorblockN{Jie Tang,~\IEEEmembership{Senior Member,~IEEE}}, and
	\IEEEauthorblockN{Wen Wu,~\IEEEmembership{Senior Member,~IEEE}} \vspace{-1cm}
\thanks{This work was supported in part by the National Natural Science Foundation of China under Grant 62201214, 62331022, 62222105, and 62201311, the Natural Science Foundation of Guangdong Province under Grant 2023A1515011753, the Fundamental Research Funds for the Central Universities under Grant 2023ZYGXZR106, the U.S. National Science Foundation under Grant CNS-2107182 and ECCS-2030029, and the Peng Cheng Laboratory Major Key Project under Grant PCL2023AS1-5 and PCL2021A09-B2. \itshape{(Corresponding author: Beixiong Zheng.)}}
\thanks{X. Xiong is with the School of Future Technology, South China University of Technology, Guangzhou 511442, and also with the Frontier Research Center, Peng Cheng Laboratory, Shenzhen 518055, China (e-mail: ftxuexiong@mail.scut.edu.cn).}
\thanks{B. Zheng and J. Tang are with the School of Microelectronics/School of Electronic and Information Engineering, South China University of Technology, Guangzhou 511442/510640, China (e-mail: bxzheng@scut.edu.cn; eejtang@scut.edu.cn).}
	\thanks{A. Lee Swindlehurst is with the Center for Pervasive Communications and Computing, University
		of California, Irvine, CA 92697 USA (e-mail: swindle@uci.edu).}
	\thanks{W. Wu is with the Frontier Research Center, Peng Cheng Laboratory, Shenzhen 518055, China (email: wuw02@pcl.ac.cn).}
	
} 
\maketitle

\begin{abstract}
Electromagnetic wave absorbing material (EWAM) plays an essential role in manufacturing stealth aircraft, which can achieve the electromagnetic stealth (ES) by reducing the strength of the signal reflected back to the radar system. 
However, the stealth performance is limited by the coating thickness, incident wave angles, and working frequencies. 
To tackle these limitations, we propose a new intelligent reflecting surface (IRS)-aided ES system where an IRS is deployed at the target to synergize with EWAM for effectively mitigating the echo signal and thus reducing the radar detection probability. 
Considering the monotonic relationship between the detection probability and the received signal-to-noise-ratio (SNR) at the radar, we formulate an optimization problem that minimizes the SNR under the reflection constraint of each IRS element, and a semi-closed-form solution is derived by using Karush-Kuhn-Tucker (KKT) conditions.  
Simulation results validate the superiority of the proposed IRS-aided ES system compared to various benchmarks.
\end{abstract}


\vspace{-0.2cm}
\section{Introduction}
Over the past few decades, electromagnetic stealth (ES) technology has attracted significant attention due to its ability to increase the difficulty of radar systems to detect and track stealth targets \cite{Pattanaik2021Astudy}.
As one of the most important techniques to achieve ES, electromagnetic wave absorbing materials (EWAMs) have been widely investigated to enhance electromagnetic wave (EW) absorption efficiency as well as broaden the absorption bandwidth \cite{ahmad2019stealth,yuan2011properties}.
Generally, existing EWAMs designs mainly focus on studying advanced materials with desired absorbing properties and multi-layer structures containing various materials, which can effectively reduce the reflected EW towards the radar by absorbing a considerable part of the incident electromagnetic energy \cite{Design2019Khan,bai2015reflections,Chen2018Plasma}. 
However, the performance of these customized EWAMs is limited by various factors, such as coating thickness, incident wave angle, and inherent material properties. 
Furthermore, with the increasing complexity of battlefield environments, rapid operational changes in the radar signal and targets with high speed can quickly change the angles of the incident waves, resulting in a further serious deterioration of stealth performance.
Therefore, there is an urgent need for developing more efficient and intelligent ES strategies to achieve satisfactory stealth performance in fast-changing environments.

Recently, innovative intelligent reflecting surface (IRS) technology has emerged as a tool for improving the performance of wireless systems \cite{Zheng2022ASurvey,Pan2022Anoverview}.
Specifically, an IRS is a planar surface composed of a large number of low-cost passive reflecting elements that can independently manipulate the amplitude and/or phase of incident electromagnetic signals. 
This enables an IRS to adjust the wireless propagation environment in response to rapid variations in the channel conditions. 
Furthermore, radio-frequency (RF) chains are not required in IRS, thereby reducing the deployment cost and energy consumption. 
This also facilitates its lightweight fabrication and compact size, opening up new possibilities for achieving improved stealth performance by incorporating IRS as a supplement to EWAMs.
Although there are many works on IRS-aided wireless communications \cite{renzo2019smart,Wu2020Towards}, only a handful focus on IRS-aided target sensing
\cite{Foundations2022Buzzi}.
While IRS are less effective than EWAMs in absorbing the incident radar energy, their amplitude and phase reconfigurability can be exploited to destructively combine with the residual EWAMs reflection to increase ES performance. This possibility has not been explored before and is the subject of this paper.

Motivated by the above, this work designs a new IRS-aided ES system that enables a target to evade radar detection in highly dynamic environments.
From the perspective of anti-radar detection, we formulate an optimization problem that minimizes the received signal-to-noise-ratio (SNR) at the adversarial radar under the modulus constraint of IRS reflection.
To address this optimization problem, we derive a semi-closed-form solution by leveraging the Karush-Kuhn-Tucker (KKT) conditions and then obtain the dual variables via solving the dual program.
Simulation results are provided to demonstrate the advantages of the proposed IRS-aided ES system compared to various benchmarks.

\vspace{-0.25cm}
\section{System Model}\label{sys}
\begin{figure}[!t]
	\centering
	\includegraphics[width=0.42\textwidth,height=0.31\textwidth]{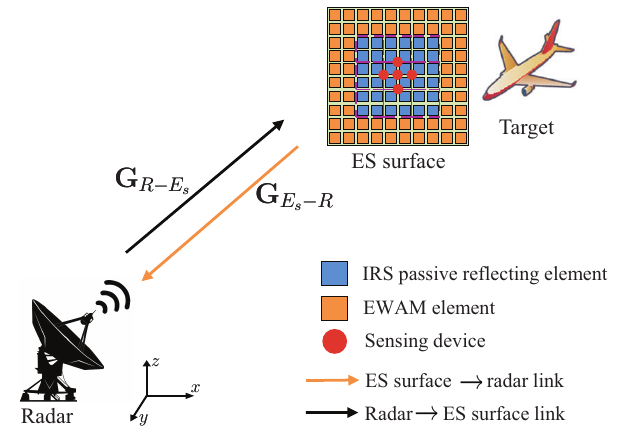}
	\setlength{\abovecaptionskip}{-0pt}
	\caption{IRS-aided electromagnetic stealth (ES) scenario.}
	\label{system}
	\vspace{-0.7cm}
\end{figure}
As shown in Fig. 1, we consider an IRS-aided ES system mounted on a moving target (e.g., aircraft), where an IRS is deployed in addition to the EWAM surface for further reducing the target detection probability. 
Furthermore, we consider a challenging scenario where a monostatic probing radar acts as an opponent that attempts to detect the target. 
Under this setting, we need to properly design the passive IRS reflection to synergize with the EWAMs, such that the probing signal echoed back from the target to the radar can be effectively mitigated or even eliminated. 
Without loss of generality, we consider a typical ES surface as a uniform planar array (UPA) that consists of $N_1$ IRS elements and $N_2 $ EWAM elements, with $N \triangleq N_{x} \times N_{z} = N_1+N_2$ total elements (see Fig. 1).
To enable the estimation of the angle-of-arrival (AoA) of the radar probing signal, a cross-shaped sensing array is embedded at the center of the ES surface, which contains  $N_s=N_{s,x}+N_{s,z}-1$ sensing devices.
Moreover, we assume that the monostatic radar system is equipped with a UPA consisting of $M \triangleq M_{x} \times M_{z}$ active antennas for either transmitting the probing signals or receiving the echo signals.

Let $\mathbf{G}_{RE_w}^{[t]} \in \mathbb{C}^{N_2\times M}$, $\mathbf{G}_{RI}^{[t]} \in \mathbb{C}^{N_1\times M}$, and $\mathbf{G}_{RE_s}^{[t]} \in \mathbb{C}^{N\times M}$ denote the equivalent baseband channels for the radar$\rightarrow$EWAM surface, radar$\rightarrow$IRS, and radar$\rightarrow$ES surface links at time $t$, respectively.\footnote{For notational convenience, we use subscripts ``$R$", ``$I$", ``$E_w$", and ``$E_s$" to indicate the radar, IRS, EWAM surface, and ES surface, respectively.} 
In a typical military scenario, considering the relatively long distance between the radar and target, the channels between them can be well characterized by the far-field line-of-sight (LoS) model.
For convenience, we first define the one-dimensional (1D) steering vector for a symmetrical uniform linear array (ULA) as follows:
\begin{equation}
	\boldsymbol{e}(\phi, N_l)=\left[e^{-j \frac{N_l -1}{2} \pi \phi }, e^{-j \frac{N_l-3}{2} \pi \phi}, \ldots, e^{j \frac{N_l-1}{2} \pi \phi}\right]^{T},
\end{equation} 
where $j = \sqrt{-1}$ denotes the imaginary unit, $\phi$ denotes the constant phase-shift difference between the signals at two adjacent antennas/elements, and $N_l$ denotes the number of antennas/elements in the ULA.
Since the IRS and EWAM surfaces are centered at the same point and oriented in the same way, they exhibit the same AoA/angle-of-departure (AoD) for the radar signals.
We denote the AoA/AoD pairs for the IRS and EWAM as $(\vartheta_{E_s}^{[t]}, \varphi _{E_s}^{[t]})$, and we denote the AoA/AoD pairs for the radar as $(\vartheta_{R}^{[t]}, \varphi_{R}^{[t]})$.
Accordingly, we let $\boldsymbol{a}_R(\vartheta_{R}^{[t]}, \varphi_{R}^{[t]})$, $\boldsymbol{a}_{E_s}(\vartheta_{E_s}^{[t]}, \varphi _{E_s}^{[t]})$, $\boldsymbol{a}_I(\vartheta_{E_s}^{[t]}, \varphi _{E_s}^{[t]})$, and  $\boldsymbol{a}_{E_w}(\vartheta_{E_s}^{[t]}, \varphi_{E_s}^{[t]})$ denote the array response vectors of the radar, ES surface, IRS, and EWAM surface at time $t$, respectively. 
Under the UPA model, each array response vector is expressed as the Kronecker product of two steering vector functions in the horizontal ($x$-axis) and vertical ($z$-axis) directions. For example, the array response vector at the ES surface can be expressed as
\begin{equation}
	\begin{aligned}
	\hspace{-0.3cm}	\boldsymbol{a}_{E_s}^{[t]}\left(\vartheta_{E_s}^{[t]}, \varphi_{E_s}^{[t]}\right)&  = \boldsymbol{e}\left(\frac{2 \Delta_{E_s}}{\lambda} \cos \left(\varphi_{E_s}^{[t]}\right) \cos \left(\vartheta_{E_s}^{[t]}\right), N_{x}\right) 
	\\ 
		\otimes & \boldsymbol{e}\left(\frac{2 \Delta_{E_s}}{\lambda} \cos \left(\varphi_{E_s}^{[t]}\right) \sin \left(\vartheta_{E_s}^{[t]}\right), N_{z}\right)
	\end{aligned}
\end{equation} 
where $\lambda$ denotes the signal wavelength and $\Delta_{E_s}$ is the element spacing at the ES surface; the array response vectors at the radar and IRS, i.e., $\boldsymbol{a}_{R}\left(\vartheta_{R}^{[t]}, \varphi_{R}^{[t]}\right)$ and $\boldsymbol{a}_{I}\left(\vartheta_{E_s}^{[t]}, \varphi_{E_s}^{[t]}\right)$ can be similarly defined. 
Due to the location relationship between the ES surface, IRS, and EWAM shown in Fig. 1, the array response vector at the ES can be decomposed into the array response vectors of the IRS and EWAM surface. 
More specifically, the relationship between these array response vectors can be described by the following equation:
\begin{equation}
	\boldsymbol{a}_{E_s}(\vartheta_{E_s}^{[t]}, \varphi_{E_s}^{[t]}) =\tilde{\boldsymbol{a}}_I(\vartheta_{E_s}^{[t]}, \varphi_{E_s}^{[t]})+ \tilde{\boldsymbol{a}}_{E_w}(\vartheta_{E_s}^{[t]}, \varphi_{E_s}^{[t]}),
\end{equation}
where \vspace{-0.1cm}
\begin{equation}
	\tilde{\boldsymbol{a}}_I(\vartheta_{E_s}^{[t]}, \varphi_{E_s}^{[t]}) = [\underbrace{0,...,0}_{{N_2}/{2}},\boldsymbol{a}_I^T(\vartheta_{E_s}^{[t]}, \varphi_{E_s}^{[t]}), \underbrace{0,...,0}_{{N_2}/{2}}]^T,
\end{equation}
\vspace{-0.1cm}
\begin{equation}
\hspace{-0.3cm}	\tilde{\boldsymbol{a}}_{E_w}\hspace{-0.1cm}(\vartheta_{E_s}^{[t]}, \varphi_{E_s}^{[t]}) \hspace{-0.1cm}=\hspace{-0.1cm}[a_1,...,a_{{\frac{N_2}{2}}},\underbrace{0,...,0}_{N_1},a_{\frac{N_2}{2}+N_1+1},...,a_{N}]^T, \vspace{-0.1cm}
\end{equation}
with $a_n$ denoting the $n$-th element of $\boldsymbol{a}_{E_s}(\vartheta_{E_s}^{[t]}, \varphi_{E_s}^{[t]})$. 
Then, the EWAM array response vector $\boldsymbol{a}_{E_w}(\vartheta_{E_s}^{[t]}, \varphi_{E_s}^{[t]})$ can be obtained by removing the zero elements of $\tilde{\boldsymbol{a}}_{E_w}(\vartheta_{E_s}^{[t]}, \varphi_{E_s}^{[t]})$.
We assume that the high-speed target moves at a constant speed of $v$, and the channel links are subjected to Doppler frequency. Accordingly, the real-time far-field LoS channels between any two nodes (represented by $X$ and $Y$ for notational simplicity) are modeled as the outer product of array response vectors at their two sides, i.e.,
\begin{equation}\label{channel}
	 \mathbf{G}_{YX}^{[t]}=\alpha_{XY}^{[t]} \boldsymbol{a}_X(\vartheta_{X}^{[t]}, \varphi _{X}^{[t]})\boldsymbol{a}_Y^T(\vartheta_Y^{[t]}, \varphi_Y^{[t]}),
\end{equation}
with $X \in\{I,E_w,E_s\}$ and $Y \in\{R\}$, where $\alpha_{XY}^{[t]}=\frac{\sqrt{\beta}}{d_{XY}^{[t]}} e^{-j2\pi(\frac{d_{XY}^{[t]}}{\lambda}+f_{XY}^{[t]}T_c)}$ is the corresponding complex-valued path gain between them at time $t$, $\beta$ denotes the reference path gain at a distance of 1 meter (m), $d_{XY}^{[t]}$ denotes the propagation distance between the two nodes at time $t$, and $f_{XY}^{[t]}=\frac{v\cos \varphi_{XY}^{[t]} \cos \vartheta_{XY}^{[t]}}{\lambda}$ is the Doppler frequency, $T_c$ is the channel coherence interval. 
For the purpose of exposition, we assume all the involved LoS channels remain approximately constant during each channel coherence time and that channel reciprocity holds for the uplink and downlink transmission. \footnote{ This assumption is practically valid since the target speed $v$ remains constant, and the marginal variations in the geometry-related parameters, i.e., distances and AoDs/AoAs, are negligible.} 
As a result, we have $\mathbf{G}_{XY}^{[t]}=\left(\mathbf{G}_{Y X}^{[t]}\right)^{T}$ and  $\alpha^{[t]}=\alpha_{XY}^{[t]}=\alpha_{YX}^{[t]}$ with $X \in\{I,E_w,E_s\}$ and $Y \in\{R\}$.

Existing EWAMs may not achieve full EW absorption due to the limitations posed by their physical characteristics, absorption frequency range, structural layout, etc. 
To characterize the stealth performance of EWAMs, we utilize the absorbing efficiency $p_n \in [0,1], n=1,...,N_2$ to represent the corresponding absorption capacity. 
To proceed, we define $\boldsymbol{\gamma}=[\gamma_1e^{j\psi_{1}}, ...,\gamma_{N_2}e^{j\psi_{N_2}}]^T \in {\mathbb{C}^{N_2 \times 1}}$ as the reflection coefficient vector of the EWAM surface, where $\gamma_n=\sqrt{1-p_n}$ and $\psi_n$ respectively represent the amplitude and phase of the $n$-th EWAM element response for $n=1,...,N_2$, which can be obtained from offline measurements. 
Furthermore, we let  $\boldsymbol{\theta}^{[t]} =\left[\beta_{1}^{[t]}e^{j\varphi_{1}^{[t]}},...,\beta_{N_1}^{[t]}e^{j\varphi_{N_1}^{[t]}} \right]^T\in {\mathbb{C}^{N_1 \times 1}}$ denote the equivalent tunable reflection coefficients of the IRS at time $t$, where $\beta_n^{[t]} \in [0,1]$ and $\varphi_n^{[t]} \in [0,2\pi)$ represent the reflection amplitude and phase shift of the $n$-th IRS element, respectively.
For the purpose of ES, the IRS reflection is devised to destructively combine with the signals reflected by the EWAM surface to reduce the reflected signal power by the entire ES surface to evade radar detection \cite{zheng2023intelligent}. 
As such, the received signals at the radar are composed of two types of echo signals reflected by the IRS and the EWAM surface:
\vspace{-0.15cm}
\begin{equation}\label{ReceivedSig}
	\hspace{-0cm}\mathbf{Y}^{[t]} =\underbrace{\mathbf{G}_{IR}^{[t]} \mathbf{\Theta}^{[t]} \mathbf{G}_{RI}^{[t]} \mathbf{S}^{[t]}}_{\text{Reflected by IRS}}+ \underbrace{\mathbf{G}_{E_wR}^{[t]} \mathbf{\Gamma} \mathbf{G}_{RE_w}^{[t]} \mathbf{S}^{[t]}}_{\text{Reflected by EWAM surface}}+\mathbf{Z}_R^{[t]},
\end{equation} 
where
$\mathbf{\Theta}^{[t]}= \mathrm{diag}\left(\boldsymbol{\theta}^{[t]}\right)$ represents the diagonal reflection matrix of the IRS at time $t$, $\mathbf{\Gamma}= \mathrm{diag}\left(\boldsymbol{\gamma}\right)$ denotes the diagonal reflection matrix of the EWAM surface, $\mathbf{S}^{[t]}=\left[\mathbf{s}_1^{[t]}, ...,\mathbf{s}_L^{[t]}\right]$ represents the transmitted radar waveform at time $t$ satisfying $\mathbb{E}\left[\mathbf{S}^{[t]}({\mathbf{S}^{[t]}})^H\right] = \mathbf{I}_{M}$ with $L > M$ representing the number of transmitted samples, and $\mathbf{Z}_R^{[t]} \in \mathbb{C}^{ M\times L}$ represents an additive white Gaussian noise (AWGN) matrix with independent elements of zero mean and variance $\sigma^2$.

\vspace*{-0.10cm}
The radar detection process during each channel coherence time can be formulated as a binary hypothesis testing problem as follows:
\vspace*{-0.1cm}
\begin{equation}\label{H}
\mathbf{Y}^{[t]}\hspace{-0.1cm}=\hspace{-0.1cm}\left\{\begin{array}{l}
		\hspace{-0.25cm}\mathcal{H}_{0}\hspace{-0.1cm}: \mathbf{Z}_R^{[t]}, \\
		\hspace{-0.25cm}\mathcal{H}_{1}\hspace{-0.15cm}: ({\mathbf{G}_{IR}^{[t]} \mathbf{\Theta}^{[t]} \mathbf{G}_{RI}^{[t]} }\hspace{-0.05cm} +\hspace{-0.05cm} {\mathbf{G}_{E_wR}^{[t]} \mathbf{\Gamma} \mathbf{G}_{RE_w}^{[t]}) \mathbf{S}^{[t]}}\hspace{-0.05cm}+ \hspace{-0.05cm}\mathbf{Z}_R^{[t]},
	\end{array}\right.
\end{equation}
where $\mathcal{H}_{0}$ and $\mathcal{H}_{1}$ represent the absence and presence of a target, respectively.
The Neyman-Pearson (NP) criterion is commonly employed for optimal decision-making in the above binary hypothesis test, i.e., $	\mathcal{T}(\mathbf{Y}^{[t]}) \underset{\mathcal{H}_{0}}{\stackrel{\mathcal{H}_{1}}{\gtrless}} \eta$, where $\mathcal{T}(\mathbf{Y}^{[t]})$ is the decision rule based on the received signal $\mathbf{Y}^{[t]}$, and $\eta$ denotes the detection threshold \cite{Richards2005fundamentals}. This criterion aims to maximize the detection probability $P_d^{[t]}$, while maintaining a predetermined false alarm probability $P_{fa}^{[t]}$ at time $t$. 
According to the NP criterion, we have
\vspace*{-0.2cm}
\begin{equation}\label{pd}
	P_{d}^{[t]}= \operatorname{Q}\left(\sqrt{2 \mathrm{SNR}^{[t]}}, \sqrt{-2 \log \left(P_{f a}^{[t]}\right)}\right),
\end{equation} 
where $\operatorname{Q}(\cdot)$ denotes the Marcum-Q function. Intuitively, a higher SNR will result in a higher probability of target detection for a given $P_{f a}^{[t]}$. Given the monotonic relationship between the SNR and detection probability in \eqref{pd}, we can reduce $P_d^{[t]}$ by decreasing the received SNR at the radar. 
Thus, we consider SNR as an indicator to evaluate the performance of the IRS-aided ES system. According to \eqref{ReceivedSig}, the received SNR at the radar can be expressed as
\begin{small}
\begin{equation}\label{SNR}
\mathrm{SNR}^{[t]}=\frac{\left \| (\mathbf{G}_{RI}^{[t]})^T \mathbf{\Theta}^{[t]} \mathbf{G}_{RI}^{[t]}+(\mathbf{G}_{RE_w}^{[t]})^T \mathbf{\Gamma} \mathbf{G}_{RE_w}^{[t]} \right \|_F^2}{\sigma^2}. \\
\end{equation}
\end{small}
\vspace{-0.7cm}
\section{Problem Formulation and Solution}
\subsection{Problem Formulation}
To achieve satisfactory ES performance, we aim to minimize the received SNR at the radar by optimizing the IRS reflection coefficients. 
Moreover, we process in a block-by-block manner, assuming that all the involved channels remain approximately constant during each channel coherence block. As such, we drop the time index $[t]$ in this section.                
Consequently, the corresponding optimization problem can be formulated as
\vspace{-0.2cm}
\begin{equation}
	\begin{array}{ll}
		\underset{\boldsymbol{\theta}}{\min} & \mathrm{SNR} \\
		\text { s.t. } 
		& \left| \boldsymbol{\theta}(n) \right| \le 1, \forall n=1,\ldots, N_1. \\
	\end{array} 
\end{equation}
By taking a closer look at the SNR expression in \eqref{SNR}, it can be simplified by exploiting the channel structure between the radar and target. In particular, by substituting \eqref{channel} into \eqref{SNR}, the SNR can be rewritten as
\vspace{-0.2cm}
\begin{equation}
	\begin{aligned}
		\mathrm{SNR}
	 = \left| \boldsymbol{a}_{I}^T \mathbf{\Theta} \boldsymbol{a}_{I} +\boldsymbol{a}_{E_w}^T \mathbf{\Gamma} \boldsymbol{a}_{E_w} \right|^2 \cdot  \frac{\left \| \alpha^2 \boldsymbol{a}_{R} \boldsymbol{a}_{R}^T\right \|_F^2}{\sigma^2},
	\end{aligned}
\end{equation}
where $\boldsymbol{a}_{R} = \boldsymbol{a}_{R} \left(\vartheta_{R}, \varphi_{R}\right)$, $\boldsymbol{a}_{I}=\boldsymbol{a}_{I}\left(\vartheta_{E_s}, \varphi_{E_s}\right)$, and $\boldsymbol{a}_{E_w}=\boldsymbol{a}_{E_w}\left(\vartheta_{E_s}, \varphi_{E_s}\right)$.
It is evident that the term $\left \| \alpha^2 \boldsymbol{a}_{R} {\boldsymbol{a}_{R}}^T\right \|_F^2/\sigma^2$ is a constant that can be omitted. 
Therefore, we only need to estimate the AoA pair $(\vartheta_{E_s},\varphi_{E_s})$ to acquire the array response vectors of both the IRS and EWAM surfaces for designing the IRS reflection.
\footnote{
The AoA information can be estimated at the cross-shaped sensing array by leveraging well-established AoA estimation algorithms such as multiple signal classification (MUSIC) algorithm \cite{Schmidt1986Multiple}.}

Accordingly, the IRS design for minimizing the overall SNR can be explicitly expressed in an equivalent form as  
\begin{equation}\label{SNR2}
	\begin{array}{ll}
		\underset{\boldsymbol{\theta} }{\min} & \left| \boldsymbol{a}_{I}^T \mathbf{\Theta} \boldsymbol{a}_{I} +\boldsymbol{a}_{E_w}^T \mathbf{\Gamma} \boldsymbol{a}_{E_w} \right|^2  \\
		\text { s.t.} 
		& \left| \boldsymbol{\theta}(n) \right| \le 1, \forall n=1, \ldots, N_1. \\
	\end{array} 
\end{equation}
{It can be verified that problem \eqref{SNR2} is a convex optimization problem and can be solved by the CVX toolbox. However, this numerical solution has a relatively high computational complexity; thus we proceed to find a low-complexity solution. 
	}

\vspace{-0.2cm}
\subsection{Solution of the Proposed Problem}
We further simply \eqref{SNR2} as follows:
\vspace{-0.1cm}
\begin{equation}\label{p1}
	\begin{array}{ll}
		\underset{{ \boldsymbol{\theta}}}{\min}  & \left |\boldsymbol{d}^H \boldsymbol{\theta} + c\right | ^2 \\
		\text { s.t.} & \boldsymbol{\theta}^H \mathbf{E} _n \boldsymbol{\theta} \leq 1, \forall n=1,\ldots, N_1,
	\end{array} 
\end{equation}
where $ \boldsymbol{d}^H =\boldsymbol{a}_{I}^T\mathrm{diag} (\boldsymbol{a}_{I})$ represents the cascaded array response at the IRS, $c= \boldsymbol{a}_{E_w}^T\mathrm{diag}(\boldsymbol{a}_{E_w})\boldsymbol{\gamma} $ is the complex reflection gain of the EWAM surface, and $\mathbf{E} _n \in \mathbb{C}^{ {N_1} \times {N_1}} $ denotes the selection matrix in which the $n$-th diagonal element is $1$, and the remaining elements are $0$. Note that the problem in \eqref{p1} is identical to one considered in \cite{Wang2023Applications} in the context of IRS absorption for radar interference mitigation.

The problem \eqref{p1} is evidently a convex quadratically constrained quadratic program (QCQP), which can be effectively solved by leveraging the KKT conditions.
Specifically, the Lagrangian function for problem \eqref{p1} can be expressed as 
%\vspace{-0.1cm}
\begin{equation} \label{Lag1}
	\mathcal{L}(\boldsymbol{\theta}, \boldsymbol{\lambda}) = (\boldsymbol{d}^H \boldsymbol{\theta} + c)^H(\boldsymbol{d}^H \boldsymbol{\theta} + c)+\sum_{n=1}^{N_1} \lambda_n (\boldsymbol{\theta}^H \mathbf{E}_n \boldsymbol{\theta}-1),  
\end{equation}
where $\boldsymbol{\lambda} =[\lambda_1,...,\lambda_{N_1}]^T $ is the dual variable vector with $\lambda_n \ge 0$, $ n = 1, ..., N_1 $. {We then take the differentiation of $\mathcal{L}(\boldsymbol{\theta}, \boldsymbol{\lambda})$ with respect to $\boldsymbol{\theta}$, denoted as $\nabla_{\boldsymbol{\theta}} \mathcal{L}$.} The corresponding KKT conditions of the primal problem thus can be derived as
\vspace{-0.2cm}  
\begin{equation}\label{dri}
\nabla_{\boldsymbol{\theta}} \mathcal{L}  = \boldsymbol{d} \boldsymbol{d}^H \boldsymbol{\theta}  + {c}  \boldsymbol{d} + \sum_{n=1}^{N_1} \lambda_n  \mathbf{E}_n \boldsymbol{\theta} = \mathbf{0},
\end{equation}
\begin{equation}
	\lambda_n (\boldsymbol{\theta}^H \mathbf{E}_n \boldsymbol{\theta}-1) =0,
\end{equation}
\begin{equation}
		\boldsymbol{\theta}^H \mathbf{E} _n \boldsymbol{\theta} \leq 1, \lambda_n \ge 0, \forall n=1, \ldots, N_1.
\end{equation}
From the condition in \eqref{dri}, a semi-closed-form solution for the optimal value of $\boldsymbol{\theta}$ can be calculated by
%\vspace{-0.1cm}
\begin{equation}\label{optimal_theta}
\boldsymbol{\theta}^{\star} = -{c}\left( \boldsymbol{d} \boldsymbol{d} ^H + \sum_{n=1}^{N_1} \lambda_n \mathbf{E}_n\right) ^{-1} \boldsymbol{d}.
\end{equation}
Obviously, the optimal value of $\boldsymbol{\theta}$ is a function of the Lagrange multiplier $\boldsymbol {\lambda}$. 
Since problem \eqref{p1} is a convex optimization problem with zero duality gap, we can obtain the dual variables $\lambda_{n} \ge 0, \forall n$ through the corresponding dual program.
According to the Lagrangian in \eqref{Lag1}, the corresponding dual function can be stated as 
\begin{equation}
\begin{aligned}
g(\boldsymbol{\lambda}) &=\underset{ {\boldsymbol{\theta}}}{\inf }  \mathcal{L}(\boldsymbol{\theta}, \boldsymbol{\lambda}) \\
		&=\left\{\begin{array}{ll}
	|c|^2-\sum_{n=1}^{N_1}\lambda_n - |c|^2 \boldsymbol{d}^{H}\mathbf{Q}^{-1} \boldsymbol{d},  \mathbf{Q} \succeq 0, \\
			-\infty, \quad \text{otherwise}, 
		\end{array}\right.
\end{aligned}
\end{equation} 
where $\mathbf{Q}= \boldsymbol{d} \boldsymbol{d}^H+ \sum_{n=1}^{N_1} \lambda_n  \mathbf{E}_n$. Using the Schur complement, we can express the dual problem as an equivalent semidefinite optimization problem:  
\vspace{-0.2cm}
\begin{equation}
	\begin{array}{cll}
		\underset{q,\boldsymbol{\lambda}}{\text{max}} & q \\
		\text { s.t. } & {\left[\begin{array}{cc}
				|c|^2-\sum_{n=1}^{N_1}\lambda_n-q & c^{\ast}\boldsymbol{d}^{H} \\
				{c} \boldsymbol{d} & \mathbf{Q} 
			\end{array}\right] \succeq 0,} \\
		& \lambda \geq 0 ,
	\end{array}
\end{equation} 
which can be efficiently solved by a standard semidefinite program (SDP) optimization package, with a complexity order of $\mathcal{O}({N_1^{4.5}})$. {\footnote{In practice, we can calculate the optimal IRS's reflection coefficients beforehand in an offline manner and store them in the IRS controller. With the precalculated database that incorporates the mapping from any given AoA information to the optimal IRS reflection coefficients, IRS can dynamically adjust the signal reflection in real time in response to the sensed AoA information.
}}

\vspace{-0.2cm}
\section{Simulation Results and Discussions}\label{Sim}
In this section, we present simulation results to demonstrate the performance of the proposed IRS-aided ES system. 
In particular, we consider two baseline systems for comparison: 1) Random phase shift design, where the IRS phase shifts are randomly generated following a uniform distribution within $[0,2\pi)$; 2) No IRS design, where the ES system works without an IRS by setting the reflection coefficient to $\boldsymbol{\theta}^{[t]}=\boldsymbol{0}$.
In our simulations, we consider that the moving target and probing radar are within the same two-dimensional plane, where the radar is located at the origin point.
Under the considered setup, we have $\varphi_{T}^{[t]} =\varphi_{R}^{[t]}=0$ for ease of illustration and only need to focus on the AoAs/AoDs $\left( \vartheta_{T}^{[t]}, \vartheta_{R}^{[t]}\right) $. 

At the target, the ES surface is composed of a total of $N=N_1+N_2 =70$ elements, with $N_1$ and $N_2$ to be specified in the following simulations. 
For the absorbing efficiency of the EWAM surface, we set $p_1=...=p_{N_2}=p$ for simplicity. Moreover, $\left \{ \psi_n \right \}_{n=1}^{N_2}$ are randomly generated following a uniform distribution within $[0,2\pi)$.
Throughout the simulations, each result is attained by independent experiments during a channel coherence time, in which the angles for each experiment are chosen randomly in the region $ [-\frac{\pi}{2},\frac{\pi}{2}]$ with a discretization stepsize of $1^\circ$.

\begin{figure}[!t]
	\centering
	\includegraphics[width=0.40\textwidth,height=0.33\textwidth]{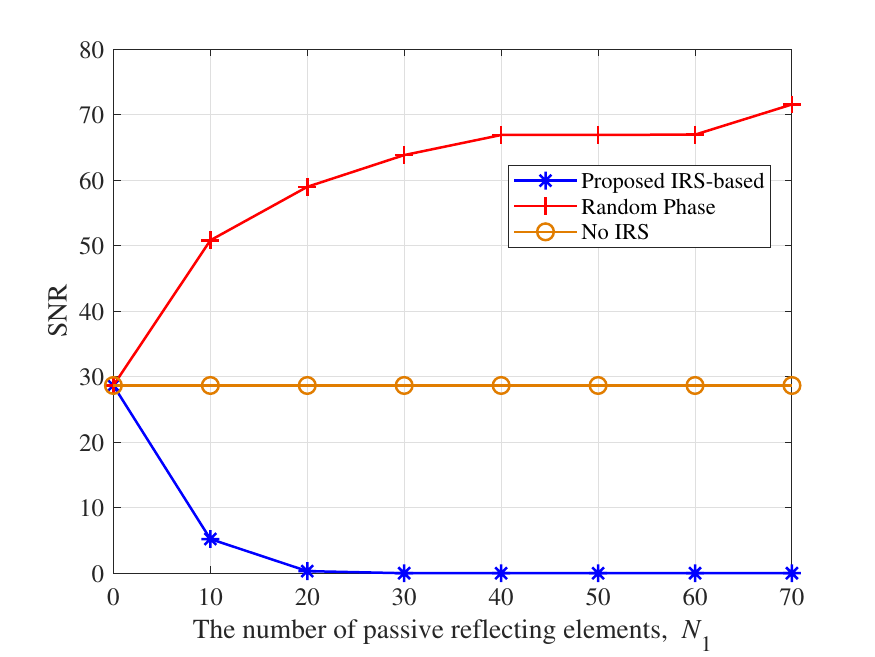}
	\setlength{\abovecaptionskip}{-0pt}
	\caption{SNR versus the number of passive reflecting elements $N_1$ with $N=70$ and $p=0.7$.}
	\label{fig2}
	\vspace{-0.5cm}
\end{figure}

In Fig. 2, we depict the SNR versus the number of passive reflecting elements for $p=0.7$. 
Several interesting observations are made as follows. 
First, as the number of IRS elements increases, the received SNR at the radar dramatically decreases due to a higher cancellation of unabsorbed signals at the EWAM surface through tuning the IRS reflection coefficients.
Second, the proposed scheme outperforms the no-IRS and random phase shift approaches in terms of the reduction in received SNR. 
This is because the signals reflected by the IRS and EWAMs are destructively combined in our proposed IRS-aided system to reduce the signal power echoed back to the radar and thus leads to a lower received SNR. 
Third, the SNR of the random phase shift design increases with the number of IRS elements and far exceeds that of the no IRS and IRS-aided designs. 
This indicates that if the signals reflected by the IRS are not designed properly to destructively combine with those of the EWAM surface, the ES performance achieved using the IRS is even worse than that without it. 
{Finally, ideal stealth effectiveness, where the SNR reaches the minimum value of $0$, can be achieved when the number of IRS elements exceeds a certain threshold, i.e., in this case $N_1\ge 20$.}

\begin{figure}[!t]
	\centering
	\includegraphics[width=0.40\textwidth,height=0.33\textwidth]{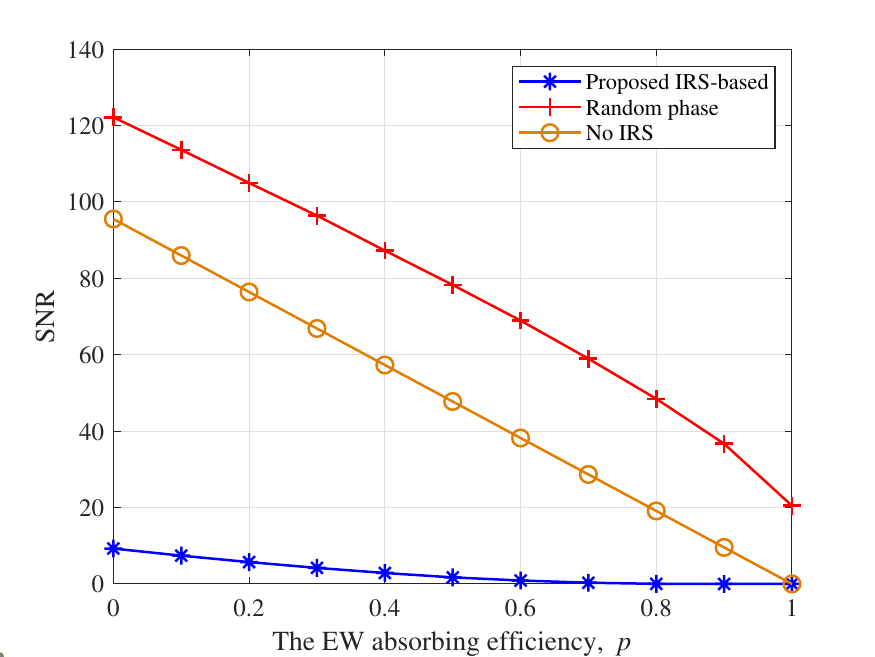}
	\setlength{\abovecaptionskip}{1pt}
	\caption{SNR versus the EW absorbing efficiency $p$ with $N=70$ and $N_1=20$.}
	\label{fig3}
	\vspace{-0.5cm}
\end{figure}
In Fig. 3, we present the received SNR versus the EW absorbing efficiency $p$ assuming an IRS with $N_1=20$ elements.
It is observed that as the EW absorption efficiency $p$ increases, the SNRs of all the considered ES systems decrease due to the reduced signal power reflected by the EWAM surface.
Moreover, the proposed IRS-aided ES system, regardless of the EW absorbing efficiency, outperforms the two baselines in terms of the reduction in received SNR at the radar.
In particular, the IRS-aided system can achieve perfect stealth performance, i.e., the received SNR at the radar reaches the minimal value of 0, even when the level of EW absorbing efficiency is not relatively high.
This is expected since the tunable elements in the ES surface can collaboratively operate with the EWAM surface for achieving better ES performance. 

\begin{figure}[htbp]
	\centering	\includegraphics[width=0.40\textwidth,height=0.33\textwidth]{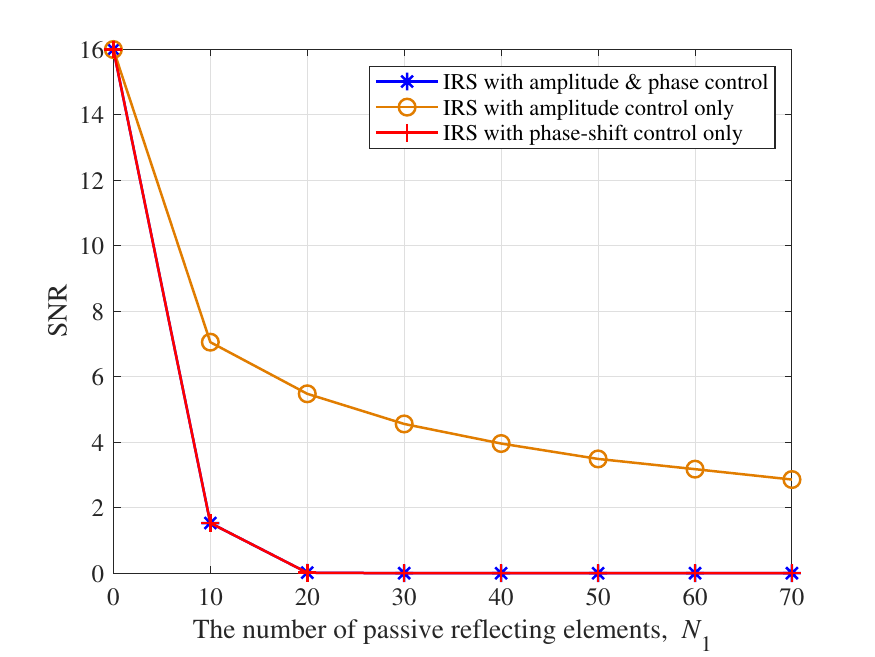}
	\setlength{\abovecaptionskip}{-1pt}
	\caption{SNR versus the number of passive reflecting elements $N_1$ with $p=0.7$ and $N_2=40$.}
	\label{fig3}
	\vspace{-0.3cm}
\end{figure}

To illustrate the impact of different reflection designs on ES performance, we consider two benchmark approaches: $1)$ IRS with amplitude control only, where each reflecting element can tune the reflection amplitude with a constant phase shift;
$2)$ IRS with phase-shift control only, where each reflecting element can tune the phase shift while the reflection amplitude is set to its maximum value of one, i.e., $|\boldsymbol{\theta}(n)|=1, \forall n=1,...,N_1$.  
In Fig. 4, we show the received SNR at the radar versus the number of passive reflecting elements at the IRS, with $N_2=40$ untunable EWAM elements.
It can be seen that the received SNR at the radar for all three IRS reflection designs decreases as the number of passive reflecting elements at the IRS increases. 
This validates the effectiveness of the proposed IRS-aided ES system for intelligently controlling the IRS to reduce or eliminate the reflected signal power towards the radar.
Moreover, an interesting observation is that, regardless of the number of IRS elements, the reflection models involving phase-shift control 
(i.e., the IRS with reflection amplitude and phase shift control, and the IRS with phase-shift control only) 
consistently yield identical optimal performance.    
They both achieve a significantly lower received SNR at the radar compared to the amplitude control only design, owing to their enhanced flexibility.
However, using the phase-control-only approach is more challenging to optimize given the non-convex constraints on the IRS element responses.

%\vspace{-0.25cm}
\section{Conclusion}
In this letter, we investigated a new IRS-aided ES system for achieving satisfactory stealth performance by reducing or even eliminating the electromagnetic waves reflected back to a radar system.
The problem was formulated as the minimization of the received SNR at the radar, subject to the modulus constraints on the IRS reflection.
By exploiting the convexity of the optimization problem, a semi-closed-form solution for optimizing the IRS reflection coefficients was derived based on the KKT conditions. 
Simulation results demonstrated that the proposed IRS-aided ES system achieves superior stealth performance compared to various baseline ES approaches.

%\section*{Acknowledgment}
%\vspace{-0.2cm}
\bibliographystyle{IEEEtran}
\bibliography{Zheng_WCL2024_0454}

\begin{thebibliography}{10}
\providecommand{\url}[1]{#1}
\csname url@samestyle\endcsname
\providecommand{\newblock}{\relax}
\providecommand{\bibinfo}[2]{#2}
\providecommand{\BIBentrySTDinterwordspacing}{\spaceskip=0pt\relax}
\providecommand{\BIBentryALTinterwordstretchfactor}{4}
\providecommand{\BIBentryALTinterwordspacing}{\spaceskip=\fontdimen2\font plus
\BIBentryALTinterwordstretchfactor\fontdimen3\font minus
  \fontdimen4\font\relax}
\providecommand{\BIBforeignlanguage}[2]{{%
\expandafter\ifx\csname l@#1\endcsname\relax
\typeout{** WARNING: IEEEtran.bst: No hyphenation pattern has been}%
\typeout{** loaded for the language `#1'. Using the pattern for}%
\typeout{** the default language instead.}%
\else
\language=\csname l@#1\endcsname
\fi
#2}}
\providecommand{\BIBdecl}{\relax}
\BIBdecl

\bibitem{Pattanaik2021Astudy}
A.~C. Binayak~Pattanaik, ``A study of stealth technology,'' \emph{Materials
  Today: Proceedings}, vol.~81, no. 2023, pp. 543--546, May 2021.

\bibitem{ahmad2019stealth}
H.~Ahmad, A.~Tariq, A.~Shehzad, M.~S. Faheem, M.~Shafiq, I.~A. Rashid,
  A.~Afzal, A.~Munir, M.~T. Riaz, H.~T. Haider \emph{et~al.}, ``Stealth
  technology: Methods and composite materials-{A} review,'' \emph{Polym.
  Compos.}, vol.~40, no.~12, pp. 4457--4472, Jun. 2019.

\bibitem{yuan2011properties}
C.-X. Yuan, Z.-X. Zhou, J.~W. Zhang, X.-L. Xiang, Y.~Feng, and H.-G. Sun,
  ``Properties of propagation of electromagnetic wave in a multilayer
  radar-absorbing structure with plasma- and radar-absorbing material,''
  \emph{IEEE Trans. Plasma Sci.}, vol.~39, no.~9, pp. 1768--1775, Sept. 2011.

\bibitem{Design2019Khan}
K.~Tayyab~Ali, L.~Jianxing, C.~Juan, R.~Muhammad~Usman, and Z.~Anxue, ``Design
  of a low scattering metasurface for stealth applications,'' \emph{Materials},
  vol.~12, no.~18, pp. 3031--3043, Aug. 2019.

\bibitem{bai2015reflections}
B.~Bai, X.~Li, J.~Xu, and Y.~Liu, ``Reflections of electromagnetic waves
  obliquely incident on a multilayer stealth structure with plasma and radar
  absorbing material,'' \emph{IEEE Trans. Plasma Sci.}, vol.~43, no.~8, pp.
  2588--2597, Aug. 2015.

\bibitem{Chen2018Plasma}
X.~Chen, F.~Shen, Y.~Liu, W.~Ai, and X.~Li, ``Study of plasma-based stable and
  ultra-wideband electromagnetic wave absorption for stealth application,''
  \emph{Plasma Sci. Technol.}, vol.~20, no.~6, pp. 065\,503--065\,513, Apr.
  2018.

\bibitem{Zheng2022ASurvey}
B.~Zheng, C.~You, W.~Mei, and R.~Zhang, ``A survey on channel estimation and
  practical passive beamforming design for intelligent reflecting surface aided
  wireless communications,'' \emph{IEEE Commun. Surveys Tuts.}, vol.~24, no.~2,
  pp. 1035--1071, Feb. 2022.

\bibitem{Pan2022Anoverview}
C.~Pan, G.~Zhou, K.~Zhi, S.~Hong, T.~Wu, Y.~Pan, H.~Ren, M.~D. Renzo,
  A.~Lee~Swindlehurst, R.~Zhang, and A.~Y. Zhang, ``An overview of signal
  processing techniques for {RIS/IRS}-aided wireless systems,'' \emph{IEEE J.
  Sel. Topics Signal Process.}, vol.~16, no.~5, pp. 883--917, Aug. 2022.

\bibitem{renzo2019smart}
M.~D. Renzo, M.~Debbah, D.-T. Phan-Huy, A.~Zappone, M.-S. Alouini, C.~Yuen,
  V.~Sciancalepore, G.~C. Alexandropoulos, J.~Hoydis, H.~Gacanin \emph{et~al.},
  ``Smart radio environments empowered by reconfigurable {AI} meta-surfaces: An
  idea whose time has come,'' \emph{EURASIP J. Wireless Commun. Netw.}, vol.
  2019, no.~1, pp. 129--148, May 2019.

\bibitem{Wu2020Towards}
Q.~Wu and R.~Zhang, ``Towards smart and reconfigurable environment: Intelligent
  reflecting surface aided wireless network,'' \emph{IEEE Commun. Mag.},
  vol.~58, no.~1, pp. 106--112, Jan. 2020.

\bibitem{Foundations2022Buzzi}
S.~Buzzi, E.~Grossi, M.~Lops, and L.~Venturino, ``Foundations of {MIMO} radar
  detection aided by reconfigurable intelligent surfaces,'' \emph{IEEE Trans.
  Signal Process.}, vol.~70, pp. 1749--1763, Mar. 2022.

\bibitem{zheng2023intelligent}
B.~Zheng, X.~Xiong, J.~Tang, and R.~Zhang, ``Intelligent reflecting
  surface-aided electromagnetic stealth against radar detection,'' \emph{arXiv
  preprint arXiv:2312.01940}, 2023.

\bibitem{Richards2005fundamentals}
M.~A. Richards, \emph{Fundamentals of radar signal processing}.\hskip 1em plus
  0.5em minus 0.4em\relax Mcgraw-hill New York, 2005.

\bibitem{Schmidt1986Multiple}
R.~Schmidt, ``Multiple emitter location and signal parameter estimation,''
  \emph{IEEE Trans. Antennas Propag.}, vol.~34, no.~3, pp. 276--280, Mar. 1986.

\bibitem{Wang2023Applications}
F.~Wang and A.~L. Swindlehurst, ``Applications of absorptive reconfigurable
  intelligent surfaces in interference mitigation and physical layer
  security,'' \emph{IEEE Trans. Wireless Commun.}, Jun. 2023, {Early Access}.

\end{thebibliography}
\vspace{12pt}

\end{document}